\documentclass[prd,twocolumn,showpacs,floatfix]{revtex4}
\usepackage{bm}
\usepackage{amssymb}
\usepackage{graphicx}
\usepackage{epstopdf}

\begin{document}
\title{Orbital evolution of a test particle around a black hole: \\ 
Indirect determination of the self force in the post Newtonian approximation}
\author{Lior M.\ Burko}
\affiliation{Department of Physics, University of Alabama in Huntsville, Huntsville,
Alabama 35899}
\date{draft of \today}
\begin{abstract}
Comparing the corrections to Kepler's law with orbital evolution under a self force, we extract the finite, already regularized part of the latter in a specific gauge. We apply this method to a quasi-circular orbit around a Schwarzschild black hole of an extreme mass ratio binary, and determine the first- and second-order conservative gravitational self force in a post Newtonian expansion. We use these results in the construction of the gravitational waveform, and revisit the question of the relative contribution of the self force and spin--orbit coupling.
\end{abstract}
 \maketitle

\section{Introduction}

The problem of finding the gravitational self force for an extreme mass ratio binary has drawn much attention over the last decade, especially because of its importance for the calculation of gravitational waveforms. The usual approach for the calculation of the self force is to calculate it {\em directly}, via a number of currently available methods \cite{poisson-lvr}. In any direct calculation, one of the major difficulties is the regularization problem of the self force, which is currently not under full control although over the last few years major new insights have been achieved towards its solution. Finding the self force for generic orbits requires such direct calculations. Once the self force is found, one needs to find the resulting orbital evolution, and to find the waveforms. Direct orbital evolution (for quasi-circuar orbits in Schwarzschild, for nonspinning \cite{paper1} and spinning particles \cite{paper2}) has been done with a scalar field self force, but hasn't been done yet for the gravitational self force. 

The most immediate problem for evolving the orbit under the gravitational self force is that we do not have as yet the latter calculated via the direct approach. An alternative approach for finding the self force is an {\em indirect} one. Specifically, one can use quantities whose effect on the orbital evolution does not require regularization---or whose regularization is already understood---and extract from them the finite, already regularized, part of the self force. A simple application of this approach is well known: One can start with the flux of  energy and angular momentum radiated to infinity (and down the horizon of the central black hole), use a balance argument to find the rate of change of the particle's constants of motion, and use the latter to find the dissipative part of the self force \cite{burko-ajp}.  This indirect determination of dissipative self force effects requires no regularization. As we show in this paper, extracting the conservative piece of the self force indirectly does require regularization, but none beyond the regularization that is needed for the derivation of the quantities that are used for the extraction of the self force.  
However, as is well known, the conservative part of the self force, which is of much practical importance for the accurate determination of gravitational waveforms, cannot be found using the flux of fields. We therefore extract it using other quantities, specifically the angular velocity. 

As the self force we derive is extracted from the angular momentum, there is no new information in it beyond that that already is encoded in the angular momentum. Therefore, the orbital evolution (and the waveforms) obtained here are not different from those obtained by other methods that are based on the angular momentum.  Specifically, in this paper we make use of a post-Newtonian expression for the angular momentum. Consequently, our waveforms are no different from post-Newtonian waveforms. The new content of this paper is in the viewpoint: the orbital evolution is obtained by a direct integration of the gravitational self force. To do this, we demonstrate how the (conservative piece of the) self force can be found from the angular momentum expression. 

In this paper, we consider the determination of the conservative piece of the self force using an indirect method---which in that sense is analogous to the determination of dissipative self forces using fluxes---and apply it to a simple situation, specifically that of a quasi-circular planar orbit around a Schwarzschild black hole, i.e., orbits that would be circular in the absence of radiation reaction. Specifically, we make use of (the post Newtonian correction of) Kepler's law to find not just the first order conservative self force, but also the second order conservative self force. After finding the self force in a specific gauge, we can use it to find orbit integrated gauge invariant quantities, such as the waveform. These results can be used for testing and checking direct results for the conservative piece of self forces, and the corresponding regularization schemes (including second order ones). We emphasize that the waveforms themselves are not new, as they are equivalent to the waveforms obtained by a post-Newtonian approach that keeps conservative effects \cite{barack-cutler}. 

This paper follows previous papers on the orbital evolution of extreme mass ratio binaries under conservative self forces. In Ref.~\cite{paper1}---henceforth paper I---we showed, using quasi-circular orbits in Schwarzschild, that conservative self forces are required for finding the waveform to $O(\epsilon^0)$, where $\epsilon$ is the mass ratio of the binary. It was also shown that second order (dissipative) self forces, i.e., self forces to $O(\epsilon^3)$, are required for  a full, self consistent determination of the waveform to this order. In Ref.~\cite{paper2}---hereafter paper II---the question of the relative importance of conservative forces and spin--orbit coupling was addressed, and in Ref.~\cite{burko-comment} the equations of motion were solved perturbatively also for spin--spin coupling, again for quasi-circular orbits. In this paper we show how one can find the conservative self force, and the resulting orbital evolution and waveform, from the post Newtonian equations of motion, not only for the first-order self force (i.e., self force to $O(\epsilon^2)$), but also the second-order self force (i.e., self force to $O(\epsilon^3)$). As pointed out in paper I, the latter are of importance for a full description of the waveform, and have gained recently much attention \cite{rosenthal}. The method of this paper, however, is not limited to the post Newtonian approach.

The organization of this paper is as follows: In Section II we recapitulate the perturbative solution of paper I for the equations of motion, and extend it to the required order. In Section III we use this solution to find the self force indirectly. In Section IV we use this self force to find the orbital evolution and the waveforms, and finally, in Section V we revisit the question of the relative importance of self forces and spin--orbit coupling.

\section{Perturbative solution for the Equations of Motion}

Let a non-spinning test body of mass $\mu$ be accelerated because of its self force 
$f_{\alpha}^{\rm SF}$ in a quasi-circular orbit around a Schwarzschild black hole of mass $M$ (with 
$\epsilon:=\mu/M\ll 1$), such that its four-velocity is $u^{\alpha}$ and its proper time is $\tau$. The equations of motion (EOM) are given by
\begin{equation}
\mu\,\frac{\,Du^{\alpha}}{\,d\tau}=f_{\beta}^{\rm SF}g^{\alpha\beta}\, ,
\end{equation}
which we solve perturbatively. 

The metric in the usual Schwarzschild
coordinates is given by
\begin{equation}
\,ds^2=-F(r)\,dt^2+F^{-1}(r)\,dr^2
+r^2\,d\Omega^2\, ,
\label{metric}
\end{equation}
where $F(r)=1-2M/r$. 
Here, $\,d\Omega^2=\, d\vartheta^2 +\sin^2\vartheta\,d\phi^2$. 

As the orbit is planar, the $\theta$ component of the EOM is trivial. 
We use the normalization condition for $u^{\alpha}$, 
namely $u^{\alpha}u_{\alpha}=-1$, to eliminate $u^t$ from the EOM 
$\,Du^{i} / \,d\tau=\mu^{-1} f_{k}^{\rm SF}g^{ik}$, $i,k=t,r,\phi$, where $D$
denotes covariant differentiation compatible with the metric
(\ref{metric}). We next use the $t$ component of the EOM  to
eliminate ${\dot u}^t$.  
We can simplify the EOM to first-order (nonlinear) ODEs by taking
$\dot{r}=V(r)$, $\dot{x}=Vx'(r)$, where $x$ denotes any quantity.
We find the EOM to be 
\begin{eqnarray}
VV'-\frac{3MV^2}{r(r-2M)}&-&(r-2M)\sigma \nonumber \\
-
\frac{1}{\mu {(u^t)}^2}\left[\left(1-\frac{2M}{r}\right)f_r^{\rm SF}
\right. 
&+& \left.\frac{V}{1-2M/r}f_t^{\rm SF}\right]=0
\end{eqnarray}
and
\begin{eqnarray}
V\sigma '&-&\frac{3MV}{r^4}+2\frac{M/r^3+\sigma}{1-2M/r}
\left[\frac{2}{r}V\left(1-\frac{3M}{r}\right)\right. \nonumber \\
&-&\left.\frac{f_t^{\rm SF}}{\mu {(u^t)}^2}\right]
-\frac{2(M/r^3+\sigma)^{1/2}}{\mu {(u^t)}^2r^2}
f_{\phi}^{\rm SF}=0\, ,
\end{eqnarray}
where ${(u^t)}^2=1/[1-3M/r-r^2\sigma-V^2/(1-2M/r)]$. 
We denote by $\Omega$ the orbital 
frequency, and by an overdot and
a prime (partial) differentiation with respect to coordinate time $t$ and
$r$, respectively. 
Here, $\sigma$ measures the deviation from
Kepler's law, i.e., 
\begin{equation}\label{om-def}
\Omega^2=M/r^3+\sigma (r)\, .
\end{equation}

This last relation is gauge independent, although each of the terms on the right hand side
are separately gauge dependent. Specifically, we fix the gauge by choosing the first term on the rhs to 
include the radial Schwarzschild coordinate $r$: under an infinitesimal coordinate transformation $x^{\alpha}\to x'^{\alpha}=x^{\alpha}+\xi^{\alpha}$ where the gauge vector $\xi^{\alpha}=O(\mu)$, the metric perturbations $h_{\alpha\beta}\to h'_{\alpha\beta}=h_{\alpha\beta}-2\nabla_{(\alpha}\xi_{\beta )}$. In the last expression, the covariant derivative operator $\nabla_{\alpha}$ is compatible with the background metric (\ref{metric}).  A gauge choice on the metric perturbations $h_{\alpha\beta}$ (e.g., the Regge--Wheeler or Lorenz gauges) is therefore equivalent to a condition on the gauge vector $\xi^{\alpha}$. The latter may have a radial component, that changes the radial coordinate description of the orbit. As noted by Detweiler \cite{detweiler}, while the angular velocity $\Omega$ is gauge invariant, the radius of the orbit $r$ is not. This means that under a gauge transformation the different terms on the rhs of Eq.~(\ref{om-def}) change, but in such a manner that their combination, or the lhs of 
(\ref{om-def}) is invariant. One may, therefore, fix the gauge (possibly up to a residual gauge freedom) by determining the ratio of the terms on the rhs of (\ref{om-def}), or, alternatively, by fixing the geometrical meaning of the symbol $r$ appearing on the rhs of (\ref{om-def}). In particular, the latter may be fixed to equal the Schwarzschild radial coordinate of the unperturbed Schwarzschild background (\ref{metric}).  By clarifying the geometrical meaning of $r$ we therefore effectively fix the gauge. Notice that this fixing is {\em not} equivalent to merely choosing which coordinates are used to describe the background geometry.

We next
expand in powers of $\epsilon$ as $\sigma (r)=\sigma^{(1)}+\sigma^{(2)}$,
$V= V^{(1)}+ V^{(2)}$, and $a_{i}= a^{(1)}_i+ a^{(2)}_i$, $x^{(j)}$
denoting the term in $x$ which is at $O(\epsilon^{j})$, and $a_i$ being
the (self) acceleration. We then expand the self force as 
$f_{i}^{\rm SF}= f^{(1)}_i+ f^{(2)}_i$, where $f^{(j)}_i=\mu a^{(j)}_i$. 
Solving perturbatively, we find the first order terms to be
\begin{equation}\label{s1}
\sigma^{(1)}=-\frac{r-3M}{\mu r^2}f^{(1)}_{r}
\end{equation}
\begin{equation}
V^{(1)}=\frac{2r}{\mu M}\frac{r-3M}{r-6M}
\left[\left(\frac{M}{r}\right)^{\frac{1}{2}}\left(1-\frac{2M}{r}\right)
f^{(1)}_{\phi}+Mf^{(1)}_{t}\right]\, ,
\end{equation}
and the second-order terms to be
\begin{eqnarray}\label{s2}
\sigma^{(2)} &=& 
-\frac{r-3M}{\mu r^2}\,\left[\,f_r^{(2)}+r^2\frac{V^{(1)}f_t^{(1)}}{(r-2M)^2}\,\right] \nonumber \\
&+&\frac{r}{\mu} \sigma^{(1)}f_r^{(1)}-\frac{3M{V^{(1)}}^{2}}{r(r-2M)^2}+\frac{V^{(1)}V^{(1)'}}{r-2M}
\end{eqnarray}
\begin{eqnarray}
V^{(2)}&=&\frac{r(r-3M)}{\mu^2M^2(r-6M)^2}
\left[2\left(\frac{M}{r}\right)^{\frac{1}{2}}f^{(1)}_{\phi}f^{(1)'}_{r}
r(r-2M)^2\right. \nonumber \\
&\times&
(r-3M)+\left(\frac{M}{r}\right)^{\frac{1}{2}}f^{(1)}_{\phi}f^{(1)}_{r}
(5r-6M)(r-2M) \nonumber \\
&\times& (r-3M)+2Mf^{(1)}_{t}f^{(1)'}_{r}r^2(r-2M)(r-3M) \nonumber \\
&+&4Mf^{(1)}_{t}f^{(1)}_{r}r^2(r-3M) +2\mu M^2f^{(2)}_{t}(r-6M)\nonumber
\\ 
&+& \left.
2\mu \left(\frac{M}{r}\right)^{\frac{3}{2}}f^{(2)}_{\phi}
(r-2M)(r-6M)\right] \, .
\label{v2}
\end{eqnarray}

In papers I and II we were primarily interested in the orbital evolution, and therefore attempted to justify neglecting the (dissipative) terms in Eq.~(\ref{v2}) that are linear in the second-order self force. In this paper we continue to assume this to be the case. Unlike the prequels, in this paper we introduce Eq.~(\ref{s2}), and leave there the (conservative) term linear in the second-order self force. In fact, we make use of Eq.~(\ref{s2}) to isolate this term, and find indirectly the (post Newtonian) expression for the second-order ($r$ component of the) self force. The first-order counterpart is found similarly from Eq.~(\ref{s1}).

\section{Finding the (post Newtonian) self force indirectly}

Consider the correction to Kepler's third law, in a post Newtonian expansion:
\begin{eqnarray}\label{omega}
\Omega^2 &=& \frac{M+\mu}{R^3} \left[ 1+(-3+\nu)\gamma + \left(6+\frac{41}{4}\nu+\nu^2\right)\gamma^2 \right.
\nonumber \\
&+& \left.
\left(-10+\rho\nu+\frac{19}{2}\nu^2+\nu^3\right)\gamma^3 +
\cdots \right]\, ,
\end{eqnarray}
where $R$ is the harmonic radial coordinate, $\nu:=M\mu/(M+\mu)^2$,  $\gamma :=(M+\mu)/R$, and $\rho$ is a certain (known)  expression \cite{blanchet-faye}. Note, that  while $\Omega^2$ is a gauge invariant quantity, it is here expressed in terms of gauge dependent quantities, specifically 
the harmonic coordinate $R$.  

This correction of Kepler's law is conservative. Equation (\ref{omega}) was derived in \cite{blanchet-faye} using Hadamard {\it Partie finie} regularization, and is correct for arbitrary mass ratios, although we make use of it here for $\nu\ll 1$. We further comment that as the self force we extract below is obtained from Eq.~(\ref{omega}), there is no new physics in the self force or the resulting orbital evolution beyond that that is already in Eq.~(\ref{omega}). 

We first recognize that the $\nu$-independent terms inside the square brackets in Eq.~(\ref{omega}) are the 
leading terms in the expansion of $(1+\gamma)^{-3}$ for $\gamma\ll 1$. Using this, we notice that 
$R^{-3}(1+\gamma)^{-3}=r^{-3}[1-3\mu/r+O(\mu^2)]$. With this substitution, we express the angular velocity as
\begin{eqnarray}\label{omega_sch}
\Omega^2 = \frac{M}{r^3} &+& \frac{\mu}{r^3} - 2\,\frac{M\mu}{r^4} + \frac{61}{4}\frac{M^2\mu}{r^5}+\cdots \nonumber \\
&-&3\frac{\mu^2}{r^4}+\frac{65}{4}\frac{M\mu^2}{r^5}+\cdots\, ,
\end{eqnarray}
where following the Keplerian term on the rhs we present in the first line of (\ref{omega_sch})
the Newtonian self force correction \cite{detweiler-poisson}, followed by the 1PN and 2PN first-order self-force corrections,
and where in the second line we present the Newtonian and 1PN second-order self-force corrections.

The correction to Kepler's law can be construed as caused by a conservative self force, that we decompose in the post Newtonian framework to the $K^{\rm th}$ PN order as
\begin{equation}
f_{i}^{(j)}=\sum_{k=0}^Kf_{i\;\;kPN}^{(j)}\, .
\end{equation}
In this decomposition we formally treat the Newtonian self force as a zeroth order post Newtonian term.

Substituting to 2PN  $\sigma^{(1)}=\mu/r^3-2M\mu/r^4+(61/4)M^2\mu/r^5$ in Eq.~(\ref{s1}), we find that 
\begin{equation}\label{fr1}
f_r^{(1)} = -\frac{\mu^2}{r^2} - \frac{\mu^2M}{r^3}-\frac{73}{4}\frac{\mu^2M^2}{r^4}+O(\mu^2M^3/r^5)\, . 
\end{equation}
This is our starting point in the investigation of the orbital evolution under the gravitational self force in the next section. 

Next, we find the radial component of the second-order self force. Substituting to 1PN $\sigma^{(2)}=-3\mu^2/r^4+(65/4)M\mu^2/r^5$ in Eq.~(\ref{s2}), we find that 
\begin{equation}\label{fr2}
f_r^{(2)}=2\frac{\mu^3}{r^3}-\frac{37}{4}\frac{\mu^3M}{r^4}+O(\mu^3M^2/r^5)\, .
\end{equation}
To this PN order, the dissipative terms in the rhs of Eq.~(\ref{s2}) do not contribute, as they enter at 2.5PN order. We can therefore ignore the terms involving dissipation, i.e., all the terms including $V^{(1)}$ or  $f_t^{(1)}$ in the extraction of $f_r^{(2)}$.

We note that the result for $f_r$, although derived from a gauge invariant quantity, certainly depends on the 
choice of gauge. For example, it was derived in Ref.~\cite{hikida} in the Regge-Wheeler gauge to 1PN. Notably, 
even the question of what is the sign of $f_r$ (and, consequently, also the question of whether it is attractive or 
repulsive) is gauge dependent. The only meaningful quantities to find, therefore, are gauge independent ones, such as 
$\Omega^2$. Indeed, when the expression of Ref.~\cite{hikida} is used to find the angular velocity, Eq.~(\ref{omega_sch}) is recovered in the appropriate PN order.

\section{Orbital evolution and waveforms}

Armed with Eq.~(\ref{fr1}) we can now integrate the orbit and find the gravitational waveforms. We use here the method of paper II with only slight modifications, and the reader is referred there for detail. The dissipative part of the self force is modeled at large distanced by the 3.5PN flux of energy to infinity 
\cite{blanchet}, and at small distances by the numerically computed luminosity in gravitational waves  
\cite{cutler}, which we fit to a polynomial to find a smooth function, from which the derivatives of $f_i^{(1)}$ can be found. As the relative error in the determination of the luminosity in either method becomes comparable at $r/M\sim 18$, we make a transition from one to the other at that distance. Specifically, we use a Fermi--Dirac type mixing, of the form 
\begin{equation}
\frac{\,dE^{\rm mix}}{\,dt}=p(r)\, \frac{\,dE^{\rm PL}}{\,dt}+[1-p(r)]\, \frac{\,dE^{\rm 3.5PN}}{\,dt}
\end{equation}
where $p(r)=1/\{\exp[(r-r_0)/\Delta]+1\}$. In practice, we took $r_0=18M$ and $\Delta=10^{-3}M$. We find this type of mixing to give better results than the sliding average we used in paper II. 

\begin{figure}[htbp]
   \centering
   \includegraphics[width=3.4in]{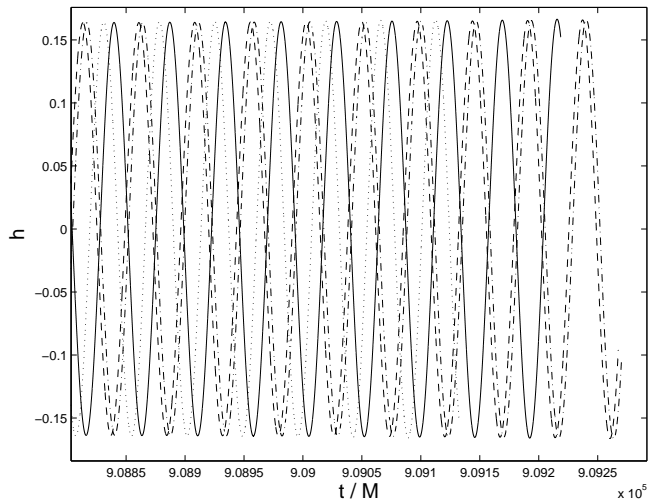} 
   \caption{The waveform for an orbit starting at $r=10M$ at $t=0$, and decaying to the ISCO at $6M$ for 
   $\mu=10^{-4}M$. Dotted line: no self force. Dash--dotted line: Newtonian self force. Dashed line: 1PN self force. Solid line: 2PN self force. }
   \label{figure1}
\end{figure}

As in papers I and II, the orbit is found through
\begin{equation}
t(r)= \int_{r_{\rm initial}}^r \frac{\, d \tilde{r}}{V(\tilde{r})}\;\;\;\;{\rm and}\;\;\;\;\varphi(r)=
\int_{r_{\rm initial}}^r \frac{\Omega(\tilde{r})\, d \tilde{r}}{V(\tilde{r})}\, ,
\end{equation}
and the waveform is obtained through the usual ``restricted waveform" approximation (see paper I for more detail). Figure \ref{figure1} shows the waveforms for an orbit that starts at $r=10M$, and decays to the inner-most stable circular orbit (ISCO) at $r=6M$  for self-force correction at the Newtonian, 1PN and 2PN orders. Not surprisingly, close to the ISCO the post Newtonian approximation breaks down.

\begin{figure}[htbp]
   \centering
   \includegraphics[width=3.4in]{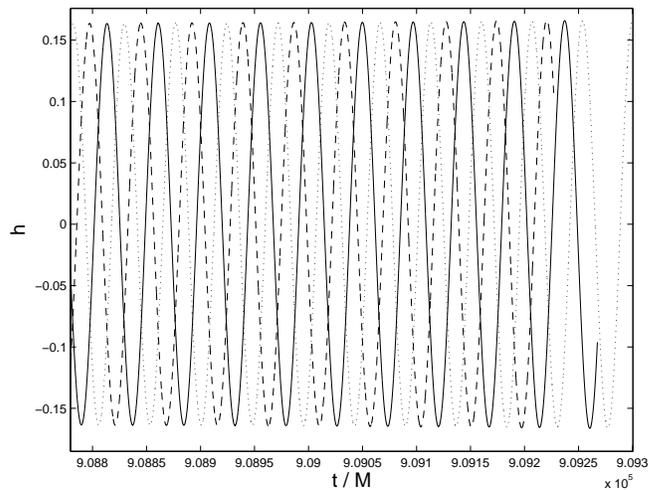} 
   \caption{The waveform for an orbit starting at $r=10M$ at $t=0$, and decaying to the ISCO at $6M$ for 
   $\mu=10^{-4}M$ at 1.5PN order for the conservative effects. Solid line: non-spinning body. Dotted line: Maximal spin aligned with the orbital angular momentum. Dashed line: Maximal spin anti-aligned with the orbital angular momentum. }
   \label{figure2}
\end{figure}

\section{On the relative contributions of self forces and spin forces}

In paper II we considered the relative importance for orbital evolution of the conservative self force and effective forces that come about because of spin--orbit coupling. The argument of paper II was that for a spinning body the spin--orbit effective force is quadratic in the body's mass, and therefore should be included in the construction of waveforms for any self-consistent description of the latter at 
$O(\epsilon^0)$. Specifically, paper II considered a conservative self force that is modeled after the scalar field self force that was found in \cite{burko-00}. Paper II then suggested that under this assumption, the self force effects on the waveforms are likely to be overwhelmed by the spin--orbit coupling effects. 

More recently, Pound {\em et al} have correctly commented, that the model adopted in paper II was irrelevant for realistic binaries, because the gravitational self force enters at a lower PN order than the scalar field counterpart \cite{pound-05}. Indeed, the conservative gravitational self force is a 1PN effect (not counting here the Newtonian self force), whereas the conservative scalar field self force is a 3PN effect \cite{burko-00}. On the other hand, the spin--orbit force is a 1.5PN effect.
It might be inferred from the argument by  Pound {\em et al} that for realistic binaries the situation is reversed: it is not the self force that can be neglected compared with the spin--orbit effect, but the other way around.

In this Section we revisit the question of the relative importance of the conservative self force and 
spin--coupling effects. We follow the method of paper II, and include the spin--orbit force in the EOM. 
Figure \ref{figure2} displays the waveform at 1.5PN order for the conservative effects for a non-spining body, and bodies with maximal spins aligned or anti-aligned with the orbital angular momentum. Bodies with intermediate values for their spin will be described by waveforms in between these two. Comparison of Fig.~\ref{figure1} and \ref{figure2} suggests that the orbit integrated spin--orbit effect is comparable to the self force effect, even though the former enters at a higher PN order than the latter. We therefore infer that any self consistent determination of the waveform to $O(\epsilon^0)$ will have to include both the self force and the spin--orbit effect.

\section*{Acknowledgments}

The author wishes to thank Eric Poisson and Steve Detweiler for discussions.


\begin{thebibliography}{99}

\bibitem{poisson-lvr} E.~Poisson, Living Rev.~Rel.~{\bf 7}, 6 (2004) and references cited therein. 
\bibitem{paper1} L. M. Burko,  Phys.~Rev.~D {\bf 67}, 084001 (2003).
\bibitem{paper2} L. M. Burko,  Phys.~Rev.~D {\bf 69}, 044011 (2004).
\bibitem{burko-ajp} L. M. Burko,  Am. J. Phys. {\bf 68}, 456 (2000). 
\bibitem{barack-cutler} L.~Barack and C.~Cutler, Phys.~Rev.~ D {\bf 69}, 082005 (2004); J.R.~ Gair, K.~Glampedakis, gr-qc/0510129. 
\bibitem{burko-comment} L. M. Burko, Class.~Quantum Grav.~{\bf 22}, S847 (2005).
\bibitem{rosenthal} E.~Rosenthal, Class.~Quantum Grav.~{\bf 22}, S859 (2005).
\bibitem{detweiler} S.~Detweiler, Class.~Quantum Gravity {\bf 22}, S681 (2005).
\bibitem{blanchet-faye} L.~Blanchet, G.~Faye, and B.~Ponsot, Phys.~Rev.~D {\bf 58}, 124002 (1998); 
L.~Blanchet and G.~Faye, Phys.~Lett.~{\bf A271}, 58 (2000);
L.~Blanchet, Living Rev.~Rel.~{\bf 5}, 3 (2002).
\bibitem{detweiler-poisson} S.~Detweiler and E.~Poisson, Phys.~Rev.~ D {\bf 69}, 084019 (2004).
\bibitem{hikida} W.~Hikida, H.~Nakano, and M.~Sasaki, Class.~Quantum Grav.~{\bf 22}, S753 (2005).
\bibitem{blanchet} L. Blanchet, in {\it Proceedings of the 16th
International Conference on General Relativity and
Gravitation}, N.T. Bishop and S.D. Maharaj (eds.) (World Scientific,
Singapore, 2002), p.p. 54-71 [gr-qc/0201050].
\bibitem{cutler} C. Cutler, L.S. Finn, E. Poisson, and G.J. Sussman,
Phys. Rev. D {\bf 47}, 1511 (1993). 
\bibitem{burko-00} L.M. Burko, Phys. Rev. Lett. {\bf 84}, 4529 (2000).
\bibitem{pound-05} A.~Pound, E.~Poisson, and B.G.~Nickel, Phys.~Rev.~D {\bf 72}, 124001 (2005). 

\end{thebibliography}
\end{document}